# The JANUS X-Ray Flash Monitor

A. D. Falcone[a], D. N. Burrows[a], S. Barthelmy[b], W. Chang[c], J. Fredley[d], M. Kelly[d], R. Klar[e], D. Palmer[f], S. Persyn[e], K. Reichard[d], P. Roming[a], E. Seifert[d], R. W. M. Smith[d], P. Wood[e], M. Zugger[d]

[a]Penn State University, Dept. of Astronomy & Astrophysics, University Park, PA 16802 USA;
[b]NASA Goddard Space Flight Center, Greenbelt, MD 20771 USA;
[c]Edge Space Systems, Glenelg, MD 21737 USA;
[d]Applied Research Lab, Penn State University, State College, PA 16804 USA;
[e]Southwest Research Institute, San Antonio, TX 78238 USA;
[f]Los Alamos National Laboratory, Los Alamos, NM 87545 USA;

**ABSTRACT**

JANUS is a NASA small explorer class mission which just completed phase A and was intended for a 2013 launch date. The primary science goals of JANUS are to use high redshift ($6<z<12$) gamma ray bursts and quasars to explore the formation history of the first stars in the early universe and to study contributions to reionization. The X-Ray Flash Monitor (XRFM) and the Near-IR Telescope (NIRT) are the two primary instruments on JANUS. XRFM has been designed to detect bright X-ray flashes (XRFs) and gamma ray bursts (GRBs) in the 1-20 keV energy band over a wide field of view (4 steradians), thus facilitating the detection of $z>6$ XRFs/GRBs, which can be further studied by other instruments. XRFM would use a coded mask aperture design with hybrid CMOS Si detectors. It would be sensitive to XRFs/GRBs with flux in excess of approximately 240 mCrab. The spacecraft is designed to rapidly slew to source positions following a GRB trigger from XRFM. XRFM instrument design parameters and science goals are presented in this paper.

**Keywords:** X-ray, CMOS, active pixel sensor, coded mask, star formation, gamma ray bursts, JANUS, XRFM

## 1. INTRODUCTION: THE MOTIVATION FOR JANUS

JANUS was selected as one of the six NASA small explorer (SMEX) class missions to proceed to the most recent phase A study period. This selection resulted from the exciting primary science goals of the mission and due to the relatively low-risk and simple mission design. These science goals include: 1) Measure the star formation rate in early Universe by detecting more than 50 GRBs at $5<z<12$, 2) Detect the brightest quasars from $6<z<10$ and measure their contribution to reionization, and 3) Study the details of reionization history and metal enrichment in early Universe. To achieve these science goals, autonomous spacecraft slewing and two primary instruments are required. The first of these instruments is the X-Ray Flash Monitor (XRFM) which will detect XRFs and GRBs, thus providing triggers for autonomous slewing of the spacecraft. The second instrument is the Near Infrared Telescope (NIRT), which is a 50 cm aperture IR telescope capable of providing GRB/XRF redshifts and enhanced positions, as well as providing a survey of distant bright quasars.

The X-Ray Flash Monitor (XRFM) is capable of facilitating these overall science goals by providing 1-20 keV X-ray monitoring with a wide field of view and with sufficient angular resolution to follow-up the XRF/GRB with narrow field instruments. All of this is provided with real time algorithms and electronics, and fast slew triggering. The soft X-ray energy band covered by XRFM is critical since it allows one to probe the spectral parameter space that would contain the peak energy from highly redshifted GRB spectra and typical XRF spectra. Additionally, XRFM will continuously monitor the X-ray sky.

## 2. THE X-RAY FLASH MONITOR DESIGN

### 2.1 Overall Instrument Design

The XRFM instrument is designed to identify and localize GRBs for follow-up by the NIRT. The instrument uses a low risk design with significant heritage from the coded aperture *Swift* BAT instrument, the *Swift* X-ray Telescope, and other missions. X-ray detectors view the sky through a random-patterned mask that casts a shadow pattern of each point source in the FoV onto the detector pixel array. In order to maximize the number of bright, high redshift bursts discovered, we use an array of X-ray detectors sensitive in the 1-20 keV band and maximize the instrument solid angle. The instrument consists of 10 identical modules (referred to as detector modules; DMs), each containing 4 identical detectors and associated Application-Specific Integrated Circuits (ASICs). Each module is pointed in a different direction on the sky to provide a FoV of 3.9 sr, which is biased in the anti-sun hemisphere to facilitate ground-based follow-up observations.

Key instrument characteristics and requirements are given in Table 1. Individual detector module (DM) sensitivity is shown in Figure 1 (right) and is background-limited. For a typical 30 s long GRB, a single DM has an on-axis sensitivity of 240 mCrabs for a 7σ detection, corresponding to a fluence limit of $2.7\times10^{-7}$ erg $cm^{-2}$ for a Crab-like spectrum. Figure 1 (left) shows the effective area curve.

**Table 1**: Key XRFM baseline (Bsln) parameters are shown, along with instrument requirements and goals that have flowed down from the science requirements.

| Parameter | Bsln | Req | Goal |
|---|---|---|---|
| Bandpass (keV) | 0.5-20 | 1 – 20 | 0.5-20 |
| Full FoV (sr) | 3.9 | 3.5 | 4 |
| Half-max FoV (sr) | 1.7 | | |
| Fully-coded FoV (sr) | 2.4 | | |
| Half-coded FoV (sr) | 2.9 | | |
| Ang Res (arcmin) | 6.3 | 10 | 5 |
| Eff. Area ($cm^2$)* | 21 | 20 | 27 |
| ΔE/E @ 5.9 keV | 3.5% | 5% | <3% |
| Time resolution | 0.17 s | 1 s | 0.1 s |
| DXRB rate* | 540 cps (1-20 keV) | | |
| Internal bkgnd | < 1 cps | | |
| GRB Sensitivity (1-20 keV)* | 240 mCrabs, 7σ, 30s ($2.7\times10^{-7}$ erg/$cm^2$) | | |
| Triggering Algorithms | Swift BAT Heritage | | |

*Single module

## 2.2 The XRFM Focal Plane Array

The heart of the XRFM is the Focal Plane Array (FPA), which consists of 10 identical DMs. A baseplate serves as the mechanical interface to the S/C bus and defines the DM pointing directions. The baseline design of the FPA is two cylindrical "caterpillar" arrays of 5 modules, with 18° separation between adjacent modules and with 40° between the midplanes of the caterpillars (Figure 2). The DM fields of view overlap significantly to provide a total FoV of 3.9 sr (Figure 3). The XRFM will be canted by 45° from the NIRT boresight, permitting overlap between the instruments but biasing the XRFM FoV towards the anti-sun direction to facilitate ground-based follow-up of GRBs discovered by JANUS.

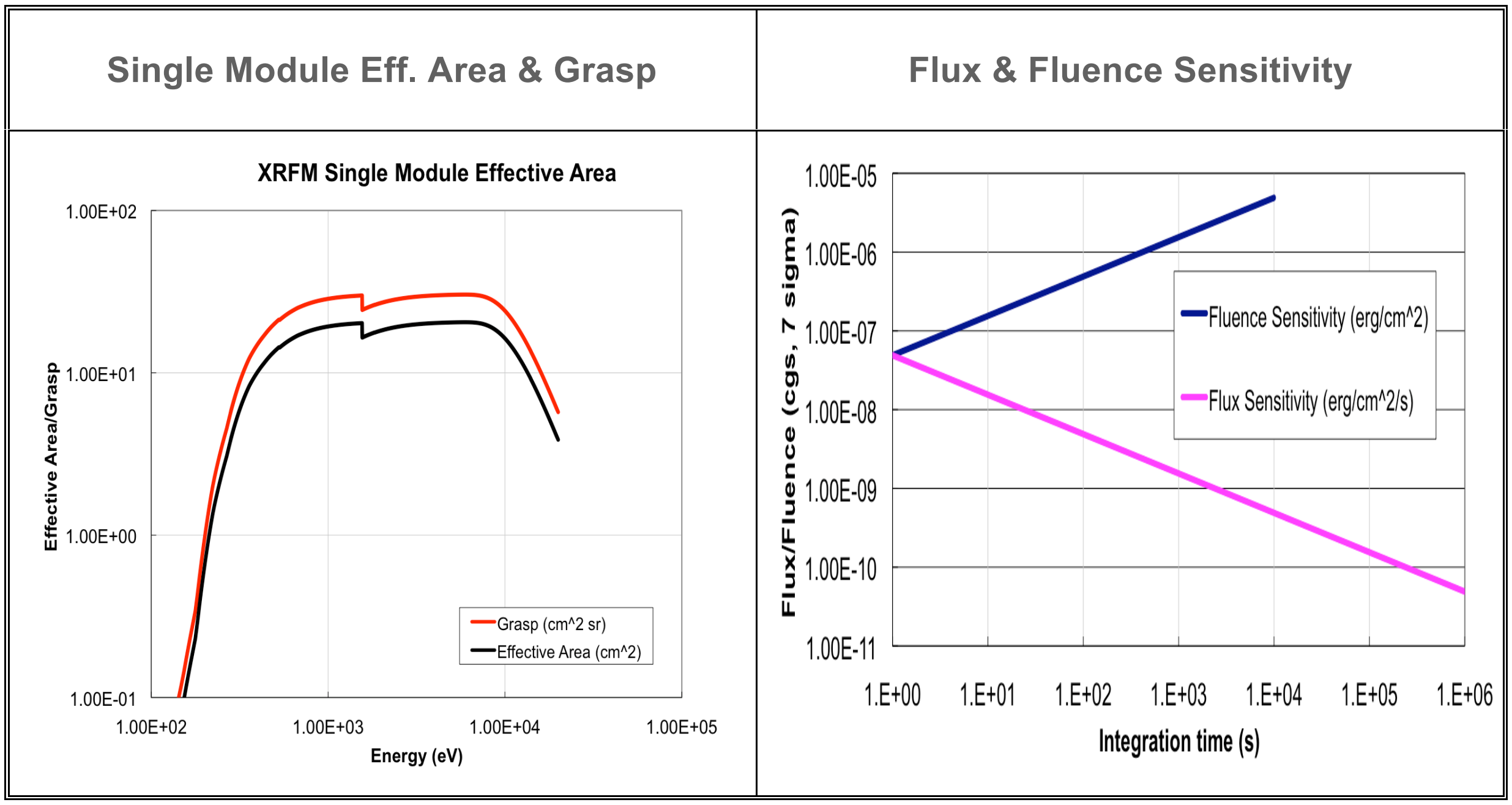

Figure 1: (left) Effective area and grasp for a single XRFM detector module, as a function of energy, (right) Flux and fluence sensitivity of a single XRFM detector module, as a function of integration time on the burst.

The modular FPA design shown in Figure 2 facilitates fabrication, testing and calibration, and prevents bright X-ray sources in the FOV of one DM from degrading the sensitivity of the entire instrument.

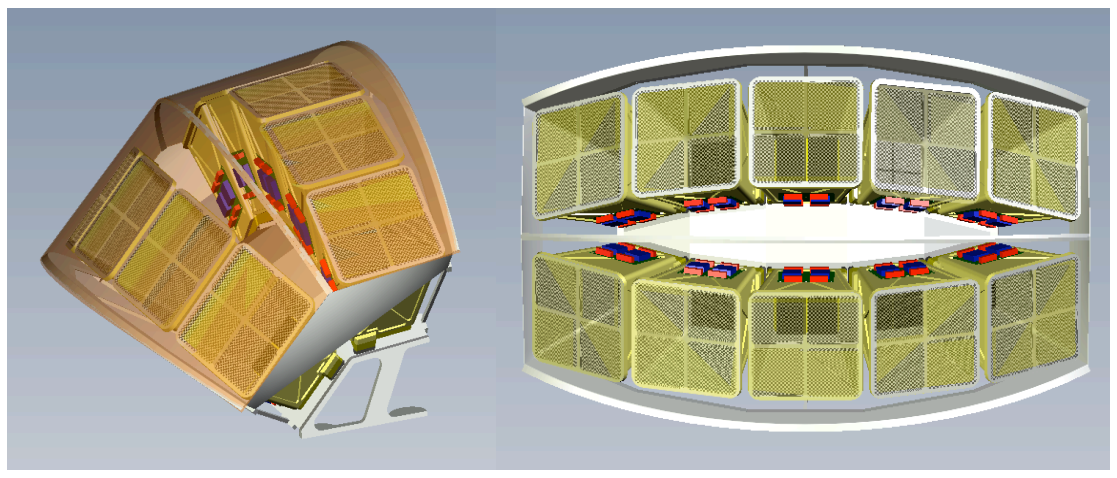

Fig. 2: XRFM Focal Plane Array. Left: side view with contamination covers in place. Right: top view with contamination covers open. A point source at the location of the viewing angle illuminates portions of 3 DMs in the top row and 3 in the bottom row.

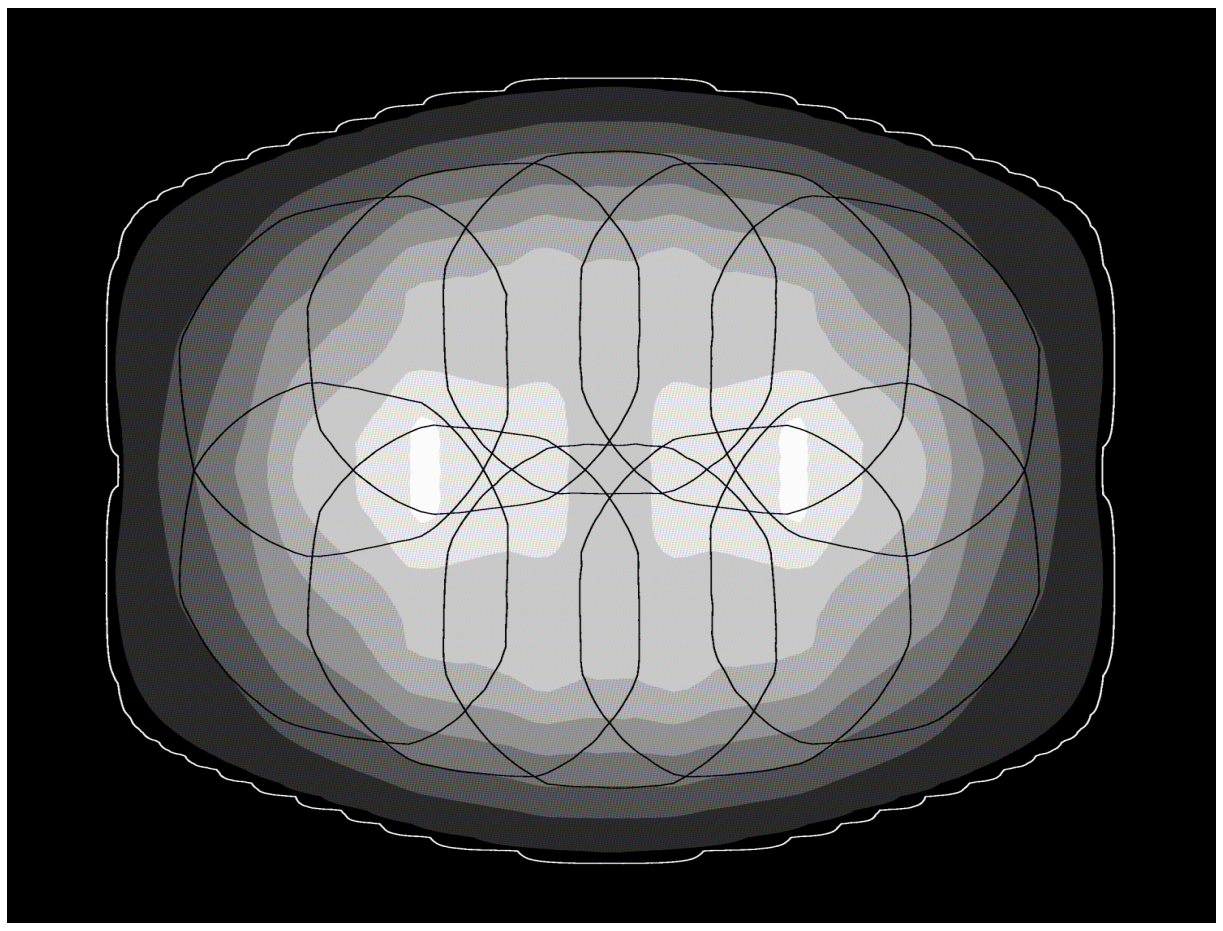

Fig. 3: XRFM field of view. The outer white contour shows the full XRFM field of view. The gray-scale shows levels of 0.001, 0.01, 0.02, 0.03, 0.04, 0.05, 0.06, and 0.07 cm2 sr. The black contours show the 50% coding contours for each individual DM.

### 2.3 The Detector Modules

The XRFM focal plane array is highly modularized. Each DM consists of a truncated inverted pyramid with four detectors in the base, a coded aperture mask to provide imaging capability, and side walls designed to absorb off-axis X-rays and to support the mask (Fig. 4). The module walls are 0.5 mm thick Invar, providing >99% absorption for X-ray energies below 20 keV from outside the FoV. The detector mosaic is a space-qualified version of the *WIRCAM* mosaic built for the CFH Telescope (Fig. 5).

### 2.4 The Detectors

The XRFM will use Teledyne HyViSI hybrid CMOS detectors [1,2,3] with H2RG readouts. These detectors were selected for their high technical readiness level, their extremely low power requirements (cf. CCDs), and the existence of a rad-hard space qualified ASIC that drives the detector and processes the output data. Detector parameters are given in Table 2.

The HyViSI detector is a hybrid CMOS detector made of a silicon photo-absorbing pixel array connected to a CMOS

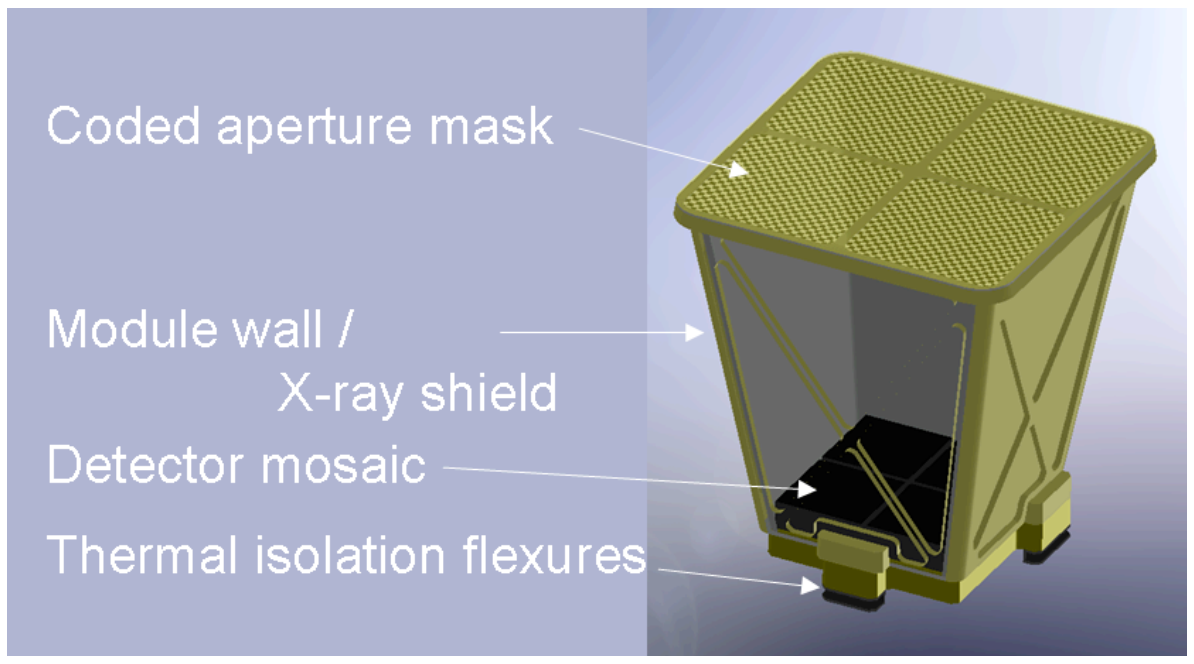

Figure 4: Single XRFM Detector Module. One side wall has been made transparent for illustrative purposes. Four detectors at the base view the sky through a random coded-aperture mask (top surface) divided into 4 panels for mechanical support.

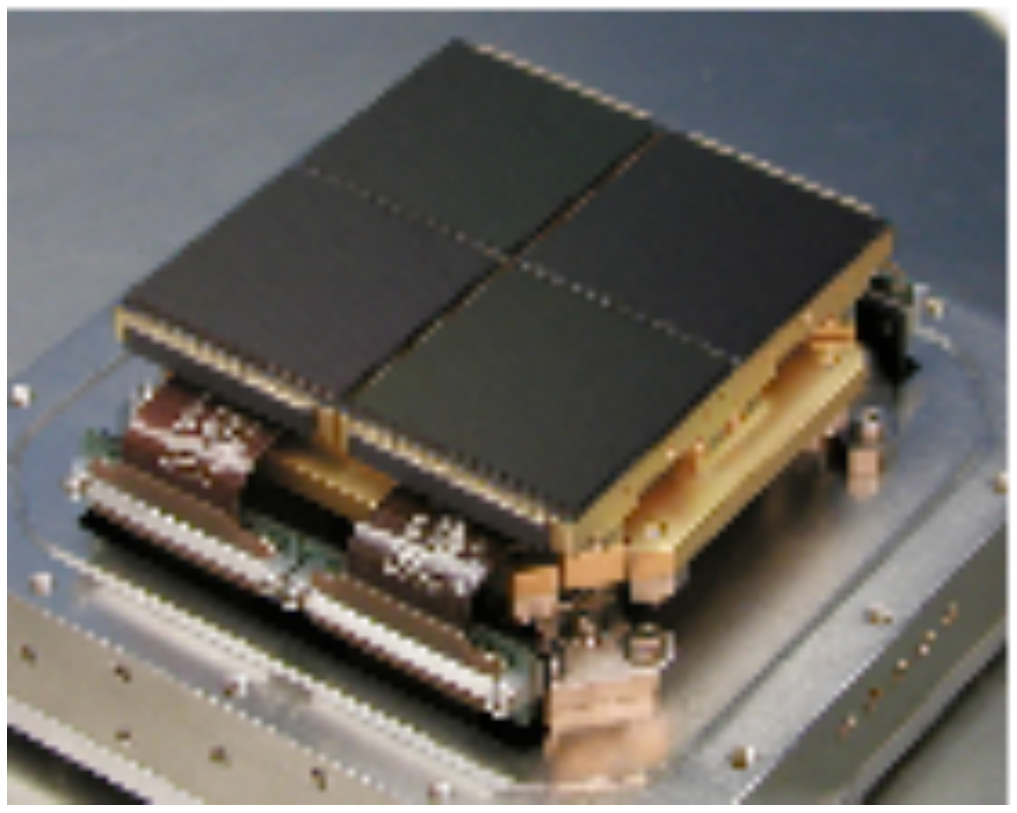

Figure 5: Existing mosaic. The XRFM will use a flight qualified version of this mosaic design.

readout multiplexer via indium bump-bonds. The native format of Teledyne's H2RG detectors uses 18 micron pixels and an optical antireflection coating. Over the past four years, Penn State University and Teledyne have worked to optimize these detectors for application to X-ray astronomy. This work has accomplished the following: replacement of optical anti-reflection filter with an Al optical blocking filter deposited directly onto the detectors; demonstration of 36 micron detector pixels hybridized to the standard 18 micron pitch readout array; optimization of the pixel design for reduced interpixel coupling, improvement in dark current, and measured energy resolution ($\Delta E/E$) of better than 5% at 5.9 keV [1]. These detectors meet the requirements of this mission.

A detector operating temperature upper limit of -40°C is set by the requirement that the total system noise in each readout be no larger than 30 $e^-$. This temperature requirement, and the associated system noise requirement, are suitable for the science goals of JANUS, and these requirements are suitable for a SMEX mission that does not have the luxury of active cooling.

**Table 2:** Detector Parameters

| Parameter | Value |
|---|---|
| Type | Hybrid CMOS |
| Absorber material | Silicon |
| Absorber thickness | 300 μm |
| Detector Format | 1024 x 1024 |
| Readout | H2RG |
| Readout rate | 200 kHz, 32 channels |
| Pixel Size | 36 x 36 μm |
| Software binning | 8 x 8 |
| Power | 10 mW |
| Operating Temp. | < -40° C |

Read noise is an increasing function of readout rate, while dark current noise is a decreasing function of readout rate, as shown in Figure 6. An optimum readout rate can be found for a given operating temperature that balances the dark current noise and the read noise. We have chosen a readout rate of 200 kilopixels per second per channel as our baseline (there are 32 output channels per detector). This gives us a frame rate of 6 frames per second and time resolution of 0.17 s. The resulting system noise is shown as a function of detector temperature in Figure 7, and meets our requirement of less than 30 $e^-$ rms system noise for T < -28C. At our maximum operating temperature of -40°C the system noise is 17 $e^-$ rms, providing ample margin. Based on thermal modeling, we expect the hot case operating temperature to be about -50°C.

Each detector is controlled by a Teledyne SIDECAR® ASIC, which provides all necessary detector control signals in addition to analog signal processing and Analog-to-Digital conversion. The XRFM SIDECAR® ASICs will operate 32 parallel channels at a pixel rate of 200 kpix/s. This ASIC replaces several boards of camera drive and signal processing electronics with a single low power integrated circuit.

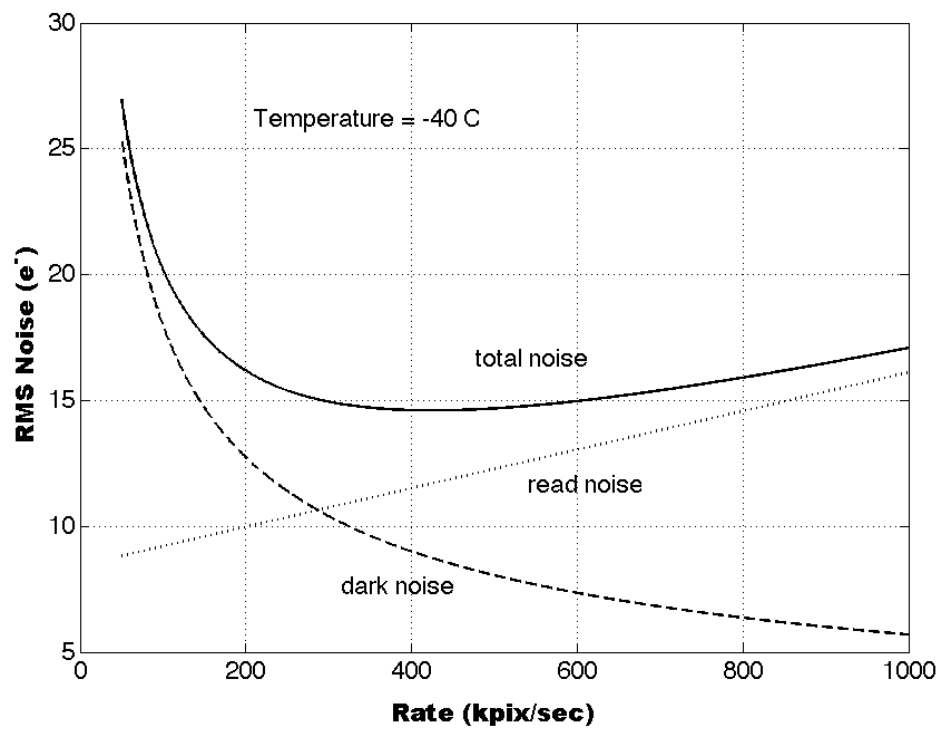

Figure 6: Estimated noise as a function of readout rate for T = -40 C. At our readout rate of 200 kpix/s per channel, we are near the minimum in total noise even a detector temperature of -40 C.

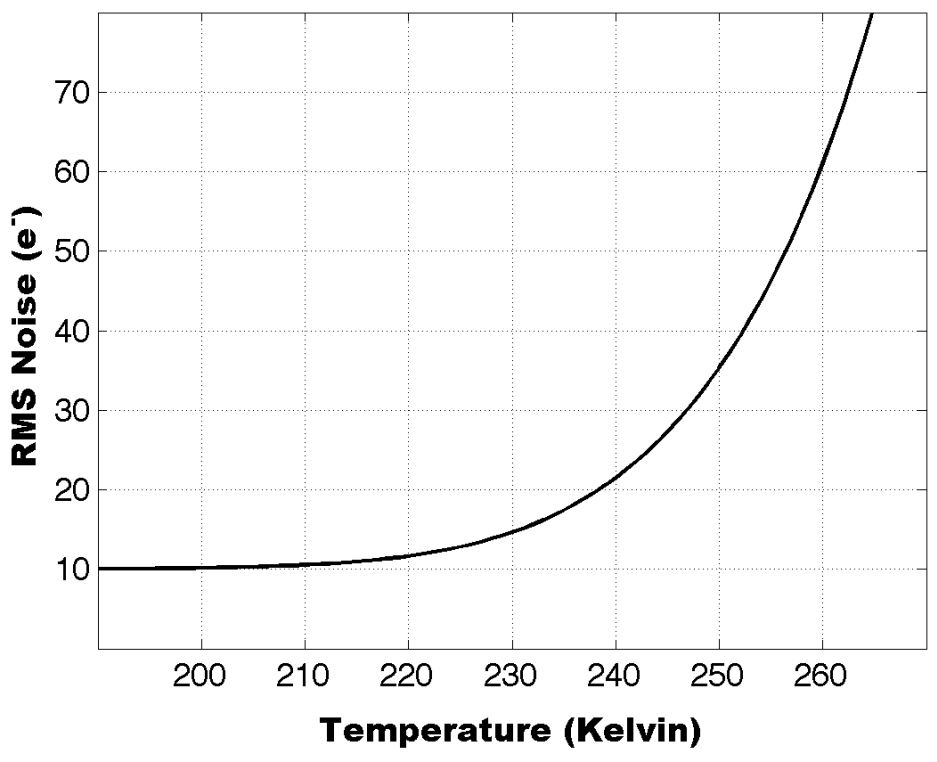

Figure 7: Total system noise as a function of temperature for a readout rate of 200 kpixels/s/channel. The total noise is below 30 e- for detector temperatures below -28 C.

## 3. XRFM THERMAL DESIGN

The XRFM FPA uses a cold-biased passive thermal design with control heaters to regulate critical temperatures. We have built a detailed 338 node thermal model of the XRFM FPA, which was incorporated into an 84-node model of the rest of the spacecraft to evaluate the thermal design. The model incorporates appropriate thermal coatings and thermal links between the XRFM detector modules and the spacecraft, based on the preliminary mechanical design. After examination of external thermal inputs and of the *JANUS* operations concept, we determined that the thermal hot case for XRFM is a $\beta=0^{o}$ orbit with the spacecraft perpendicular to the Earth-Sun line during orbit day and pointed toward the local zenith during orbit night. The cold case is a $\beta=50^{o}$ orbit with similar orientations. The XRFM detectors are passively cooled utilizing a radiator on the top plate of the spacecraft. The detector and mask temperatures are actively regulated under software control by heaters to better than 1° C over orbital time scales. The mean hot case detector temperature is -52.6 C, and the mean cold case temperature is -76.0°C. The mask temperature is held at +20°C. These simulations have shown that the XRFM detectors and module components can be held to the required temperatures with passive cooling and active heaters.

## 4. XRFM ELECTRONICS DESIGN

We have completed an end-to-end electronics board design and verified that the XRFM on-board data processing requirements can be met at a frame rate of 6 frames per second. Figure 8 shows a high-level block diagram of the electronics design. Design studies showed that we needed to separate the signal processing functions of event recognition and the subsequent processing (including FFT) into two boards. A Front-end Processor Board (FPB) accepts data from the DMs at a rate of 256 million pixels per second over 1280 parallel data channels (multiplexed onto 40 LVDS pairs with the four detectors from each DM processed by one FPGA), extracts and "grades" X-ray events based on their pattern on the detector, and passes up to 50,000 events per second to the Auxiliary Processing Board (APB) for further processing.

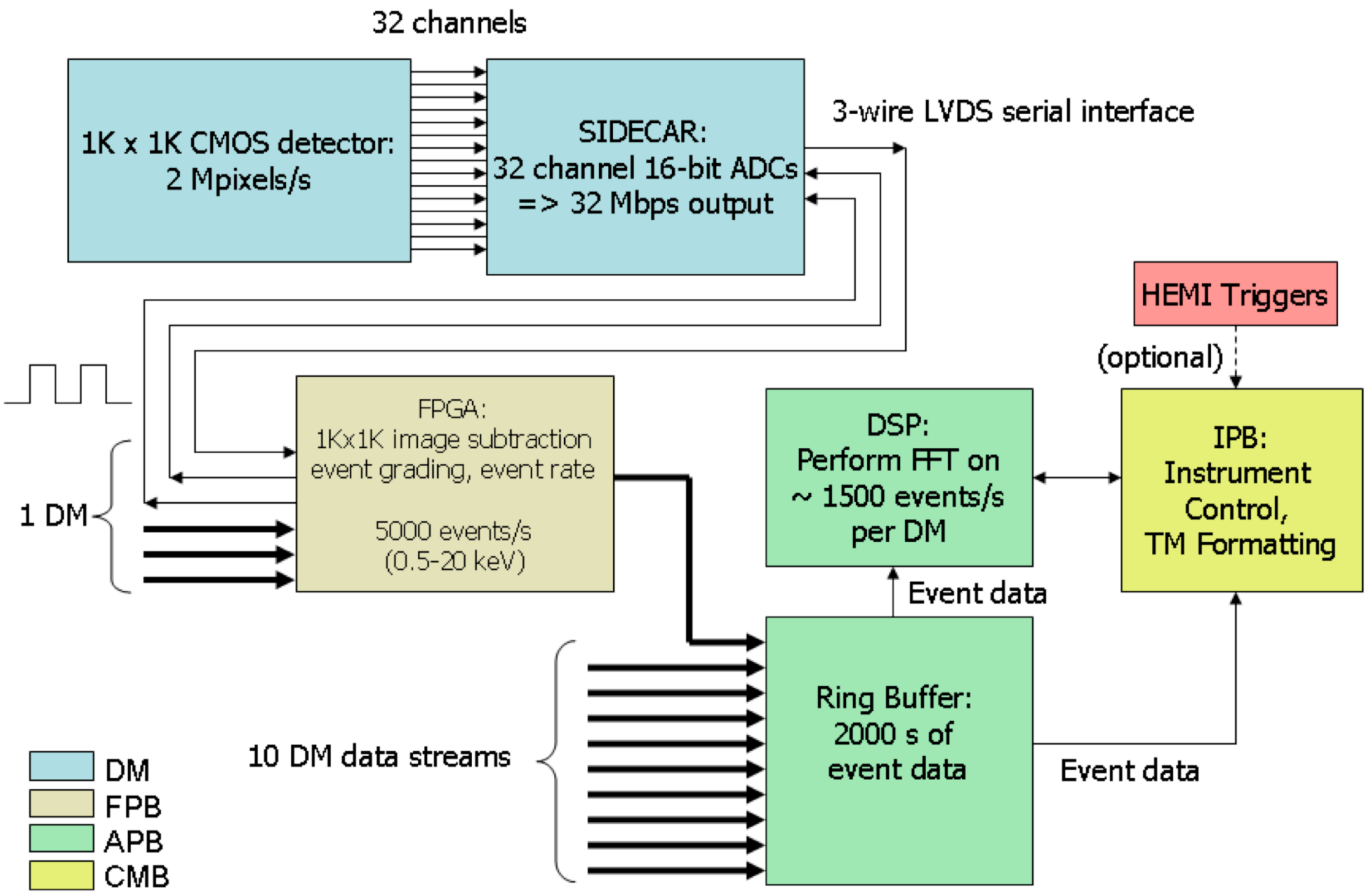

Figure 8: Block diagram of the on-board data processing flow

The APB accepts a stream of X-ray events from the FPB containing detector coordinates, energy, event grade, and time. The APB bins detector pixels into 8 x 8 superpixels, constructs event rate and detector plane histograms, performs Fast

Fourier Transforms (FFTs) on the X-ray event maps, and deconvolves them from the mask pattern. Using this method, the board calculates the positions of bright X-ray sources on the sky and sends these positions to the instrument processor board (IPB). It also provides a list of scrubbed X-ray events to the IPB. The heart of this board is a Texas Instruments SMJ320C6701, a 32-bit digital signal processor (DSP) which is rad-hard (100 krad and no latchup to 89MeV) and Class V space rated. Estimates of processing speed show that the XRFM will be able to determine GRB positions within 5 seconds after the burst.

In addition to the FPB and the APB, described above, the XRFM electronics box houses 4 other boards. The instrument processor board (IPB) incorporates a space-qualified SPARC8 processor. The communications/memory board (CMB) supports communication with the spacecraft via redundant SpaceWire links for both telecommands and telemetry, and it includes a master instrument clock. The housekeeping/relay board (HKRB) collects, scales, and digitizes housekeeping information to pass to the CMB, and controls solid state relays that actuate the XRFM heaters. The power supply board (PSB) accepts unconditioned spacecraft power at 28 ± 4V and supplies regulated, filtered DC power at all voltages required by the instrument. These six boards are housed within the XRFM Electronics Package (XEP), which consists of a single enclosure with a design with heritage from the Deep Impact Flyby Spacecraft.

## 5. XRFM FLIGHT SOFTWARE

XRFM software will borrow significantly from the heritage of Swift, which has similar instrument system software, bootstrap, command & data-handling (C&DH) infrastructure, and triggering algorithms. The science software is normally in survey mode. X-ray events are collected from 40 detectors, graded to eliminate cosmic rays, and placed into a ring buffer (sized to store ~2000 seconds of data) by the FPGAs on the FPB. Next, the science software searches the buffered data using the same sophisticated triggering algorithms employed by *Swift* BAT. Energy histograms are accumulated for each detector superpixel, while the software simultaneously searches event data and count rates for GRB triggers in multiple energy bands, spatial detector regions, and time intervals. Spikes in count rates, or *rate triggers,* are verified by calculating a sky image (using the DSP), and checking them against a recent background image for evidence of a variable source to effectively eliminate false triggers.

The DSP performs fast Fourier transforms (FFTs) in the background while sky images are searched by the main processor for slowly varying sources. Triggers found in this way are called *image triggers*, and have been used by *Swift* BAT to discover many interesting GRBs including GRB/SN 060218 and the high-z GRB 050904.

When a trigger is detected, the software transitions to Trigger Mode. The data in the event buffer are frozen and trigger data products are generated. If a GRB meets the observing criteria, a message is sent to the spacecraft control unit (SCU) requesting a slew (for observation by the NIRT). Trigger data products include a GRB centroid position, a total count-rate light curve, and a detector pixel map. Survey data products include 8 energy-channel histograms for each 8x8-binned superpixel plus rate and energy histograms.

## 6. XRFM DATA PRODUCTS

XRFM data products for GRB triggers include a GRB centroid position, a total count-rate light curve, and a detector pixel map. These will be rapidly relayed to the ground through TDRSS to support evaluation of the event and to facilitate ground-based follow-up.

GRB event data will include a 2000 second interval centered on the trigger, tagging each event with position, energy, and time (using 256 energy channels and full spatial resolution). These data will be stored in a high-priority solid state recorder partition and will be transferred to the ground preferentially during ground contacts. This will require about 60 Mbytes/GRB.

Sky survey data will be accumulated in 6 energy channel histograms for each 8x8-binned superpixel. Sky survey data will be transferred to a low priority recorder partition at the end of each NIRT observation during slew to the next source. These detector histograms require 190MBytes/day. Additional data products include count rate data (light curves), energy histograms, and housekeeping data, with a combined data rate of about 20 MBytes/day.

## 7. CALIBRATION

The XRFM detectors will be tested at Teledyne Imaging Systems before delivery to Pennsylvania State University (PSU) and will be calibrated at X-ray energies at PSU following acceptance tests. The 2x2 detector mosaic calibrations will focus on measurements of gain and energy scale, dark current, bad pixels, quantum efficiency and energy resolution as a function of energy. These measurements will be performed at multiple temperatures spanning the anticipated on-orbit operating temperatures. Effective area will be measured to 10% accuracy relative to a calibrated detector in the same beam.

Following mosaic calibration, each DM will be assembled and integrated into the FPA. When acoustic testing is complete the entire FPA will be installed in the 45m long X-ray Calibration Facility at PSU and will be operated by our Engineering Model electronics box. A Manson source provides calibration energies between 0.3 and 12 keV. This full end-to-end instrument test/calibration will measure mask alignments and angular offsets of the array in addition to providing additional energy calibration data using the full on-board data processing. Bursts will be simulated to verify proper instrument response to transient X-ray point sources (using radioactive sources with controllable shutters).

The DMs will incorporate weak $^{55}$Fe sources designed to allow us to monitor on-orbit calibration changes over time intervals of several days. On-orbit effective area measurements will be made using the Crab Nebula as a reference source. Detector alignment will also be measured on-orbit with respect to the star tracker/attitude control system axes.

## 8. DISCUSSION AND CONCLUSIONS

By utilizing past flight heritage whenever possible and by choosing a modular designed coded-aperture telescope, we have shown that XRFM can achieve all of the requirements imposed upon it by the JANUS mission. This was achieved within the confines of a tight NASA SMEX budget. The JANUS mission was not selected to proceed to phase B in the current NASA SMEX cycle. If the XRFM were completed, it would be capable of detecting highly redshifted GRBs and XRFs, thus providing a means to study the early universe. XRFM would also be capable of monitoring the entire X-ray sky in the 0.5-20 keV band, thus contributing to the study of a multitude of transient and variable astrophysical sources.

## 9. ACKNOWLEDGEMENTS

We acknowledge financial support from NASA contract NNG08FC02C and from The Pennsylvania State University.